\documentclass[12pt]{article}
\usepackage{a4}

\usepackage{amsfonts}
\usepackage{amsmath}
\usepackage{amssymb}

\makeatletter

\@addtoreset{equation}{section}
\makeatother

\begin{document}
\allowdisplaybreaks{
\begin{titlepage}
\begin{center}

\hfill 
\vskip 2.8cm
{\Large \bf Quantum corrections of (fuzzy) spacetimes  } \\[3mm]
{\Large \bf from a supersymmetric reduced model}\\[3mm]
{\Large \bf with Filippov 3-algebra}

\vskip 22mm 
{\sc Dan Tomino}\footnote{e-mail address: tomino@phys.cts.nthu.edu.tw} \ \ \ 
\vskip 10mm
{\sl
National Center for Theoretical Sciences\\
National Tsing-Hua University, Hsinchu 30013, Taiwan, R.O.C.}\\
\noindent{ \smallskip }\\

\vspace{8pt}
\end{center}

\begin{abstract}
1-loop vacuum energies of (fuzzy) spacetimes from a supersymmetric reduced model with Filippov 3-algebra are discussed. 
$A_{2,2}$ algebra,  Nambu-Poisson algebra in flat spacetime, and a Lorentzian 3-algebra are examined as 3-algebras.  
\end{abstract}
\end{titlepage}

\renewcommand{\thefootnote}{\arabic{footnote}}
\setcounter{footnote}{0}

\section{Introduction}
Gauge symmetry based on Filippov 3-algebra (or, Lie 3-algebra) \cite{Filippov}  has been applied in the study of M-theory in recent years. It is used to write down an effective theory of multiple M2-branes ending on the M5-brane (the BLG model) \cite{Bagger:2007jr}. Other M-theory objects such as the M5-brane are also obtained from the BLG model if  one particularly choose Nambu-Poisson algebra as a 3-algebra \cite{Ho:2008nn}. 
Recently, the use of Nambu-Poisson algebra to produce the KK monopole from the BLG model was proposed \cite{Huang:2010yz}, 
and an on-shell supersymmetry algebra of the non-Abelian (2,0) tensor multiplet in six dimension was written down using the 3-algebra in \cite{Lambert:2010wm}.
The Nambu-Poisson algebraic structure also naturally appears in a toy model of membrane field theory \cite{Ho:2007vk}.

On the other hand, there is a different approach of using  3-algebra as a tool to study  multiple membranes. 
It is  known that the Green-Schwarz type supermembrane with lightcone gauge can be regularized using large size matrices and the resulting action is supersymmetric Yang-Mills matrix quantum mechanics \cite{de Wit:1988ig}
\!\!\footnote{
Afterward, a large $N$ limit of this super-Yang-Mills quantum mechanics was reinterpreted as a formulation of M-theory in the infinite momentum limit (BFSS model) \cite{Banks:1996vh}.}. 
However, if we choose a different gauge to fix the kappa-gauge symmetry, 
we can use the  3-bracket of  Lie 3-algebra to "regularize" the membrane world volume.
 Because we do not take the lightcone gauge which partially breaks the Lorentz invariance of the target spacetime,  the resulting models manifestly retain  full Lorentz invariance. 
The supersymmetric reduced models with Lie 3-algebra structure in \cite{Furuuchi:2009ax} were constructed on the basis of such an idea, 
and these are expected to be useful toy models for investigation of the fundamental natures of multiple M2-brane
\!\footnote{
In these models, numbers of dimension of target spacetimes are less than eleven. Therefore, they should be considered as toy models of the  M2-brane. 
Supersymmetric reduced models related to a  membrane in eleven dimension are discussed 
in \cite{Hanada:2009hd, Sato:2010ca}, but they do not have complete eleven-dimensional Lorentz invariance.}. \\

There are many similarities between these reduced models in \cite{Furuuchi:2009ax} and the IKKT matrix model \cite{Ishibashi:1996xs}. 
The IKKT matrix model is a supersymmetric Yang-Mills  reduced matrix models with gauge symmetry of Lie 2-algebra, and it relates to the Green-Schwarz  I{}IB superstring through a matrix regularization of the string world sheet. 
From this point of view, the reduced models with  3-algebraic structure are a kind of generalization of the IKKT model
\!\footnote{
Such  generalization of the  BFSS model was presented in \cite{Lee:2010ey}. 
A suggestion to use $p+1$-algebra for $p$-branes was made in \cite{Kamani:2009wg}  
}. 
Therefore, to investigate the  dynamics of these 3-algebraic models, it seems to be natural to carry out analysis similar to that done for the IKKT model,

In IKKT type matrix model, the matrix background describes an extended object and the matrix model around this background yields a gauge theory on a non-commutative space described by the matrix background (see \cite{Connes:1997cr, Aoki:1999vr, Iso:2001mg, Kimura:2003ab, Kitazawa:2002xj, Ishii:2008tm, Kawai:2009vb} for examples). 
The quantum correction of the non-commutative space is calculated in terms of the non-commutative gauge theory, and we can discuss quantum stabilities of these non-commutative space. 
The main motivation of this paper is to follow such analysis for the case of a reduced model with Lie 3-algebra, according to the similarity between these Lie 2- and Lie 3-algebraic models. 

In this paper, we expand the reduced model with 3-algebra around a background. 
In general, the meaning of such background is less clear than the background of the matrix model, but we may have a gauge theory on a fuzzy spacetime nevertheless.   
  After suitable gauge fixing it is possible to calculate the quantum correction through an analogue of the path integral. 
We then carry out a preliminary study of such quantum correction: investigating of 1-loop determinants around several particular backgrounds. 
These determinants can be interpreted as 1-loop vacuum energies of each fuzzy spacetime. 
The paper \cite{Furuuchi:2009ax} discussed reduced models with $n$-algebraic structure $n=3,4$ and target space dimension $D=4,5,6$. 
In the present paper, however,  we only consider the minimal case of $n=3,D=4$ for simplicity. \\

The outline of the paper is as follows. In section 2, we review the supersymmetric reduced model with Lie 3-algebraic structure and state the subject of this paper to study. 
We then consider 1-loop determinants for several  choices of 3-algebra. $A_{2,2}$ algebra is considered in section 3, 
a Nambu-Poisson algebra  in section 4, and the simplest Lorentzian 3-algebra  in section 5. Section 6 presents a summary and discussion.


\section{Set Up}
\subsection{Reduced Model}
To write the reduced model, we use the 3-algebraic structure
\begin{eqnarray}
[T^a,T^b,T^c]={f^{abc}}_dT^d 
\end{eqnarray} 
($a,b,c$ are  totally anti-symmetric) 
 which satisfies the property of the so-called fundamental identity:
\begin{eqnarray}
[T^a, T^b, [T^c, T^d, T^e]]&=& 
[[T^a, T^b, T^c], T^d, T^e] +[T^c,[ T^a, T^b, T^d], T^e]
\nonumber \\ &&
+[T^c, T^d, [T^a, T^b, T^e]],
\label{fumdid}
\end{eqnarray}
where $T^a$ is generator of the 3-algebra with inner product
\begin{eqnarray}
\langle T^a T^b \rangle=h^{ab}.
\label{inpro}
\end{eqnarray}
The inverse of $h^{ab}$ is denoted as $h_{ab}$. We also impose invariance of the metric $h^{ab}$:
\begin{eqnarray}
\langle [T^a,T^b,T^c]T^d \rangle+\langle T^c[T^a,T^b,T^d] \rangle=0.
\label{metinv}
\end{eqnarray} 
Imposing the condition of  (\ref{metinv}), $f^{abcd}={f^{abc}}_eh^{ed}$ becomes totally anti-symmetric. 

Using these structures, the reduced model action is written as  
\begin{eqnarray}
S=-\frac{1}{12}\left\langle [\hat{X}^I, \hat{X}^J, \hat{X}^K][\hat{X}_I, \hat{X}_J, \hat{X}_K]\right\rangle
+\frac{1}{4}\left\langle \bar{\Psi}\Gamma^{IJ}[\hat{X}_I,\hat{X}_J,\Psi] \right\rangle, 
\label{action}
\end{eqnarray} 
where $\Phi=\Phi_aT^a$,  $\Phi=(\hat{X}^I, \Psi)$. 
$\hat{X}^I \;(I=1,2,3,4)$ is a boson, $\Gamma^{IJ}=\frac{1}{2}\Gamma^{[I}\Gamma^{J]}$,
and   $\Gamma^I$ is  a $SO(2,2)$ Gamma matrix. $\Psi$ is a Majorana-Weyl fermion with a projection condition $\Gamma_5\Psi=-\Psi$. 
The reason for this rather unusual choice of the $SO(2,2)$ Gamma matrix  is that 
the Majorana-Weyl fermion does not exist in four-dimensional ordinary Minkowski spacetime with $SO(3,1)$ Lorentz symmetry and Euclidean space with $SO(4)$ Lorentz symmetry. Explicit representations of $SO(2,2)$ Gamma matrices are summarized in the appendix.

Symmetries of the reduced model are as follows. First, the action is invariant under an infinitesimal 3-algebraic gauge transformation: 
\begin{eqnarray}
\delta_{\Lambda}\Phi=\sum_{a,b}\Lambda_{ab}[T^a,T^b,\Phi].
\label{gaugetransform}
\end{eqnarray} 
The second symmetry of the action is a fermionic symmetry:
\begin{eqnarray}
\delta_{\epsilon}\hat{X}^{I}=i\bar{\epsilon}\Gamma^I\Psi, \quad
\delta_{\epsilon}\Psi=\frac{i}{6}[ \hat{X}_I,\hat{X}_J,\hat{X}_K]\Gamma^{IJK}\epsilon
\label{fersymm}
\end{eqnarray} 
 where $\epsilon$ is a Majorana-Weyl fermion whose projection condition is $\Gamma_5\epsilon=\epsilon$, and $\Gamma^{IJK}=\frac{1}{3!} \Gamma^{[I}\Gamma^J\Gamma^{K]}$. 
To have  fermionic symmetry (\ref{fersymm}), $\Psi$ and $\epsilon$ must be Majorana-Weyl or pseudo Majorana-Weyl fermions.
Third and fourth symmetries are two shift symmetries: 
\begin{eqnarray}
\delta_{\xi} \hat{X}^{Ia}=\delta^{a\odot}\xi, \quad \delta_{\xi}\Psi=0,  \label{shift1}\\
\delta_{\zeta} \hat{X}^{Ia}=0, \quad \delta_{\zeta}\Psi^a=\delta^{a\odot}\zeta.  \label{shift2}
\end{eqnarray} 
where the  symbol $\odot$ indicates the center. The last symmetry is the $SO(2,2)$ Lorentz symmetry. 
It is discussed in  \cite{Furuuchi:2009ax} that 
if we identify (\ref{shift1}) with the spacetime translation then
a combination of these symmetry algebras forms the ${\cal N}=1$ super Poincar\'{e} algebra of the four-dimensional spacetime, up to the gauge transformation (\ref{gaugetransform}) and equations of motion.  \\  

\subsection{Quantum Correction: 1-Loop Determinant}
Quantum theory of the reduced model may be defined by an analogue of the path integral. 
Therefore we identify  the integral 
\begin{eqnarray}
Z=\int {\cal D}\hat{X}{\cal D}\Psi\;e^{iS}
\label{partitionfunc}
\end{eqnarray} 
as the partition function. We now consider to carrying out this path integral around some background of $\hat{X}^I$, say $p^{\mu}$. Taking the decomposition 
\begin{eqnarray}
\hat{X}^{\mu}=p^{\mu}+X^{\mu} \quad (\mu=1, 2,..., d), \quad 
\hat{X}^i=\phi^i  \quad (i=d+1, ..., D=4), 
\end{eqnarray}
where $X^I, \phi^1$,  and fermion $\Psi$ are identified with fluctuation around $p^{\mu}$, integrating over $X^{\mu}, \phi^i$ and $\Psi$ gives the  quantum correction of the background. 
We expand the action (\ref{action}) using these fluctuations and obtain    
\begin{eqnarray}
&&S=S^{(0)}+S^{(1)}+S^{(2)}+O(p^3), \label{S}\\
S^{(0)}&=&-\frac{1}{12}\left\langle [p^{\mu}, p^{\nu}, p^{\rho}][p_{\mu}, p_{\nu}, p_{\rho}]\right\rangle, \\
S^{(1)}&=&\frac{1}{2}\left\langle X^{\mu}[p^{\nu},p^{\rho},[p_{\mu},p_{\nu},p_{\rho}]]\right\rangle,
\label{S1}\\
S^{(2)}&=& \hspace{-3mm}
\left\langle
\frac{1}{4}X^{\mu}(
P^2{\delta^{\mu}}_{\nu}+2P^{\mu\rho}P_{\rho\nu}-4[P_{\mu\rho},P^{\rho\nu}])X^{\nu}
\!+\!\frac{1}{4}\phi^iP^2\phi_i
\!+\!\frac{1}{4}\bar{\Psi}\Gamma^{\mu\nu}P_{\mu\nu}\Psi \!\!\right\rangle \!.
\label{S2}
\end{eqnarray}  
Here $S^{(1)}$ is a tadpole term that vanishes if the background $p^{\mu}$ satisfies the equation of motion 
\begin{eqnarray}
[p^{\nu},p^{\rho},[p_{\nu},p_{\rho},p_{\mu}]]=0.
\end{eqnarray} 
In $S^{(2)}$, we defined $P^{\mu\nu}$ as
\begin{eqnarray}
P^{\mu\nu} \bullet =[p^{\mu},p^{\nu},\bullet],
\end{eqnarray}
and $P^2=P^{\mu\nu}P_{\mu\nu}$, and  $[P^{\mu\rho},P_{\rho\nu}]= P^{\mu\rho}P_{\rho\nu}-P_{\rho\nu}P^{\mu\rho}$. Note that we used (\ref{metinv}) to obtain (\ref{S1}) and (\ref{S2}).  
The $S^{(2)}$ is the free Gaussian part and  $S^{(p>2)}$ are identified with interaction terms. 
Although the Leibniz rule is not required for $P^{\mu\nu}$ in general, it seems to be natural to think of $P^2$ as an analogue of the D'Alembertian $\partial^{\mu}\partial_{\mu}$ 
.

In the presence of background $p^{\mu}$, the parameter of the gauge transformation (\ref{gaugetransform}) can be expanded as
\begin{eqnarray}
\Lambda_{ab}=\kappa(p) h_{ab}+c_1{p^{\mu}}_{\!\!a}\tau_{\mu b}(p)+c_2{p^{\mu}}_{\!\!b}\tau_{\mu a}(p)
\label{backgauge}
\end{eqnarray} 
Substituting this expression into (\ref{gaugetransform}), the first term on the left-hand side  of  (\ref{backgauge}) vanishes owing to the  anti-symmetry of $f^{abcd},$ 
and the  anti-symmetric combination of the second and third terms survives. 
Therefore, in the presence of a background, the gauge transformation (\ref{gaugetransform}) becomes
\begin{eqnarray}
\delta_{\lambda} X^{\mu}&=&P^{\mu\nu}\lambda_{\nu}+[p^{\nu},\lambda_{\nu} ,X^{\mu}], 
\label{bggaugetransf1}\\
\delta_{\lambda} \phi^i&=&[p^{\mu},\lambda_{\mu},\phi^i], 
\label{bggaugetransf2} \\
\delta_{\lambda} \Psi^i&=&[p^{\mu},\lambda_{\mu},\Psi^i].
\label{bggaugetransf3}
\end{eqnarray}

The 1-loop determinant is the simplest quantum correction of the reduced model  calculated by the $S^{(2)}$ with a suitable gauge fixing. 
Here we adapt BRST gauge fixing and discuss the calculation of the 1-loop determinant. 
The BRST transformation for original action (\ref{action}) can be introduced as 
\begin{eqnarray}
\hat{\delta}_{B}{\hat X}_a^I=C_{ab}{\hat X}^{I}_b, \quad 
\hat{\delta}_{B}C_{ab}=C_{ab}C_{bc}
\end{eqnarray}  
where $C_{ab}$ is  the FP ghost corresponding to the gauge parameter $\tilde{\Lambda}_{ab}=f_{abcd}\Lambda^{cd}$. 
In the presence of the background $p^{\mu}$, 
we formally introduce a new ghost ${c_{\mu}}^a$ that is similar  to 
(\ref{backgauge})
\begin{eqnarray}
C_{ab}={f_{abd}}^e{c_{\mu}}^d{p^{\mu}}_e.
\end{eqnarray}         
 Then the BRST transformation of $X^{\mu}$ becomes
\begin{eqnarray}
\delta_{B} X^{\mu}&=&P^{\mu\nu}c_{\nu}+[p^{\nu},c_{\nu} ,X^{\mu}].
\end{eqnarray} 
\!\footnote{Transformation of $c_a^{\mu}$ is defined formally through $\hat{\delta}_B C_{ab}$.} 
We also introduce the anti-ghost $\bar{c}_a^{\mu}$ corresponding to $c_a^{\mu}$ and the  Nakanishi-Lautrup (NL) field $B_{a}^{\mu}$. 
They transform as
\begin{eqnarray}
\delta_{B}\bar{c}_{\mu}=B_{\mu}, \quad \delta_{B}B_{\mu}=0.
\end{eqnarray}
We now deform the action by adding a BRST-exact term with a gauge parameter $\alpha$: 
\begin{eqnarray}
\delta_B\left[-\bar{c}_{\mu}( \alpha B^{\mu}+P^{\mu\nu}X_{\nu} )\right]
=-B^{\mu}(\alpha B_{\mu}+P^{\mu\nu}X_{\nu})+\bar{c}_{\mu}P^{\mu\nu}\delta_B X_{\nu}.
\end{eqnarray}
 Integrating our NL field $B_{\mu}$ gives a gauge fixing term $S_{gf}$ and a ghost Lagrangian $S_{gh}$ in the form
\begin{eqnarray}
S_{gf}&=&\frac{1}{4\alpha}\left\langle(P^{\mu\nu}X_{\nu})(P_{\mu\rho}X^{\rho})\right\rangle, \\
S_{gh}&=&\left\langle\bar{c}_{\mu}P^{\mu\nu}P_{\nu\rho}c^{\rho}
+\bar{c}_{\nu}P^{\mu\nu}[p^{\rho},c_{\rho},X_{\nu}]\right\rangle.
\end{eqnarray}
Taking $\alpha=\frac{1}{2}$, the quadratic part of the gauge fixing action becomes
\begin{eqnarray}
&&S^{(2)}+S_{gf}+S_{gh}^{(2)} \rightarrow \nonumber \\
&&\hspace{-5mm}\left\langle
\frac{1}{4}X^{\mu}(
P^2{\delta^{\mu}}_{\nu}-4[P_{\mu\rho},P^{\rho\nu}])X^{\nu}
+\bar{c}_{\mu}P^{\mu\nu}P_{\nu\rho}c^{\rho}
+\frac{1}{4}\phi^iP^2\phi_i
+\frac{1}{4}\bar{\Psi}\Gamma^{\mu\nu}P_{\mu\nu}\Psi \right\rangle. 
\label{gaugefixedS2}
\end{eqnarray}
We now formally use the Gauss-Fresnel integral of $X^{\mu},c^{\mu}, \bar{c}^{\mu},\phi^i$ and $\Psi$, and thus $(\ref{gaugefixedS2})$ gives the 1-loop determinant expressed as
\begin{eqnarray}
\frac{ 
\det_{c\bar{c}} (iP_{\mu\nu}P^{\nu\rho})
\det^{1/2}_{\Psi} (iC^{-1}\Gamma^{IJ}P_{IJ} ) }
{\det_{X}^{1/2}( i(P^2{\delta^{\mu}}_{\nu}-4[P_{\mu\rho},P^{\rho\nu}]))
\det_\phi^{1/2} (iP^2)}
\label{1-loop}
\end{eqnarray}       
Unfortunately, this formal expression is not  so useful or accurate. One reason is that, as we will see later,  
the gauge transformation properties of fluctuations can differ depending on the structure of the 3-algebra. 
We need more detailed information on each 3-algebra to clarify this.  \\

From the next section, we choose particular examples of 3-algebra and study 1-loop determinants in each case to make a primary observation of this problem.


\section{$A_{2,2}$ Algebra}
$A_{2,2}$ algebra is a Lorentzian version of $A_{4}$ algebra. It is given by
\begin{eqnarray}
[p^I, p^J, p^K]=i\epsilon^{IJKL}p_L, 
\label{A22}
\end{eqnarray}
where $I,J,K,L$ run to 1,2,3,4 with signature $\eta_{IJ}=(-1,-1,+1,+1)$.    
In this section, we use $A_{2,2}$ algebra as the 3-bracket in the reduced model action. At the same time, we also think that (\ref{A22}) gives a background.  Therefore, we take the decomposition $\hat{X}^{I}=p^I+X^I$, and fluctuations $X^I$ and $\Psi$ around $p^I$ are expanded as 
\begin{eqnarray}
X^I=\sum_{J=1}^4X^{IJ}p_J, \quad \Psi=\sum_{J=1}^4\Psi_Jp^J.
\end{eqnarray}
$A_{4}$ algebra appears to describe $S^3$, while $A_{2,2}$ algebra describes a hyperbolic space $H^{2,2}$ ($AdS_3$) \cite{Chu:2008qv,DeBellis:2010pf}. Because there are only four generators to expand fluctuations, this background can be considered to  describe a small fuzzy $AdS_3$
\!\footnote {
If there is a 3-algebra that has $A_{4}$ (or $A_{2,2}$) as a sub-algebra, it can be used to describe larger size fuzzy $S^3$ (or $AdS_3$). Such 3-algebra has not yet been found \cite{Chu:2010nj}.}.

It is now convenient to decompose $X^{IJ}$ as  
\begin{eqnarray}
     T&\equiv&{X^I}_I, \label{Tra}\\
S^{IJ}&\equiv&\frac{1}{2}(X^{IJ}+X^{JI})-\frac{1}{4}\eta^{IJ}T \label{Sym},  \\ 
A^{IJ}&\equiv&\frac{1}{2}(X^{IJ}-X^{JI}) \label{Ansym} 
\end{eqnarray}
where $T$ is the trace part, $S^{IJ}$ is symmetric traceless, and $A^{IJ}$ is the  anti-symmetric part of $X^{IJ}$.
 From (\ref{bggaugetransf1})-(\ref{bggaugetransf3}) gauge transformations for $T,S^{IJ},A^{IJ}$  and $\Psi^I$ are written as
\begin{eqnarray}
\delta_{\lambda} T &=& -i\epsilon^{ABCD}\lambda_{AB}A_{CD}, \\ 
\delta_{\lambda} S^{IJ}&=& \frac{i}{2}\epsilon^{ABC(J}\lambda_{AB}{X^{I)}}_{C}
+\frac{i}{4}\eta^{IJ}\epsilon^{ABCD}\lambda_{AB}A_{CD}  , \\ 
\delta_{\lambda} A^{IJ} &=& \frac{i}{2}\epsilon^{IABJ}\lambda_{AB}+\frac{i}{2}\epsilon^{ABC[J}\lambda_{AB}{X^{I]}}_{C}, \\ 
\delta_{\lambda} \Psi^I &=& i\epsilon^{ABCI}\lambda_{AB}\Psi_{C}. 
\end{eqnarray}
According to  definitions (\ref{Tra})-(\ref{Ansym}), the bosonic part of the action $S^{(1)}+S^{(2)}$ is
\begin{eqnarray}
3T-\frac{1}{2}S_{IJ}S^{IJ}+\frac{3}{2}A_{IJ}A^{IJ}-\frac{15}{8}T^2.
\end{eqnarray}   
Here we have a tadpole term for $T$ because (\ref{A22}) is not a solution of the  equation of motion. 
Furthermore, we observe tachyonic eigenvalues of $P^2$  because the relative sign
 differs for the   $S_{IJ}S^{IJ}$ term and $A_{IJ}A^{IJ}$.
Hence, we consider  a fuzzy $AdS_3$, and  the existence of these tachyonic modes might not be problem
\!\footnote{ However, all tachyonic modes already exceed the Breitenlohner-Freedman bound $-1<m^2R^2$ for continuum $AdS_3$ with a radius $R=1$. Therefore, there could be  instabilities due to these tachyonic modes.}.
However, there seems to be another problem: there are too many negative norm states.  
We recall that $p^1$ and $p^2$ represent time-like directions, so  eight tensor components
$A_{13}, A_{14}, A_{23}, A_{24}, S_{13}, S_{14}, S_{23}, S_{24}$
correspond to the negative norm state in continuum theory. 
Because there are too few degrees of freedom of the gauge transformation to remove all of these modes, one suspects that this is not a sensible model.  

These problems are improved if we deform the reduced model action by a fourth- order term:  
\begin{eqnarray}
S\rightarrow S-\frac{i}{8} \epsilon_{IJKL} \left\langle\hat{X}^{I}
[\hat{X}^{J}, \hat{X}^{K}, \hat{X}^{L}] \right\rangle.
\label{deformedaction}
\end{eqnarray}
Now (\ref{A22}) solves the equation of motion of the deformed reduced model:
\begin{eqnarray}
[p^J,p^K,[p_I,p_J,p_K]]-i\epsilon_{IJKL}[p^J,p^K,p^L]=0.
\end{eqnarray} 
The bosonic part of $S^{(1)}+S^{(2)}$ with the deformation term  becomes
\begin{eqnarray}
S_{IJ}S^{IJ}+\frac{9}{4}T^2.
\end{eqnarray}        
At this time, there is no tadpole term and no tachyonic instability. Absence of the $A_{IJ}A^{IJ}$ term is consistent with the gauge transformation properties of $T$ and $A^{IJ}$.
Next, we  use gauge degrees of freedom $\lambda_{AB}$ to eliminate quadratic terms of $S_{13},S_{14}, S_{23},S_{24}$ (corresponding to negative norm states) and
$S_{12}, S_{34}$ (corresponding to positive norm states)
\!\footnote{This procedure can be done by introducing BRST ghost terms. In this case, we do not have quadratic terms of ghosts according to the form of $\delta_{\lambda} S^{IJ}$.}. As a result, we have three bosons with $P^2_{S}=4$ and one boson with $P^2_T=\frac{3}{2}$, where eigenvalues of $P^2$ are identified from coefficients of the quadratic terms.  

We  next consider the fermionic part of $S^{(2)}$. To calculate the fermionic  determinant, let us consider the eigenvalue equation
\begin{eqnarray}
\frac{1}{2}C^{-1}\Gamma^{IJ}[p_I,p_J,\Psi]=E\Psi.
\label{evofpsi}
\end{eqnarray} 
Using (\ref{evofpsi}) twice,  we have 
\begin{eqnarray}
3\Psi-2(\Gamma^{IJ}\Psi_Ip_J)=E^2\Psi,
\end{eqnarray}
or   
\begin{eqnarray}
2\Gamma^{IJ}\Psi_J=(E^2-3)\Psi^I.
\label{evofpsi2}
\end{eqnarray}
Again using (\ref{evofpsi2})  twice, we have  
\begin{eqnarray}
12\Psi^I+8\Gamma^{IJ}\Psi_J=(E^2-3)^2\Psi^I.
\label{evofpsi3}
\end{eqnarray}
From (\ref{evofpsi2}) and (\ref{evofpsi3}), we have the equation 
\begin{eqnarray}
(E^2-3)^2-4(E^2-3)=12.
\end{eqnarray}
 Thus, four fermions have $E^2=1$ and  another four have $E^2=9$.  
 We can now perform the Gauss-Fresnel integral using  $S^{(2)}$ with the deformation term (and gauge fixing). The 1-loop determinant then becomes 
\begin{eqnarray}
\frac{ \left(E^2_{(=9)}i\right)^{4/4} \left(E^2_{(=1)}i\right)^{4/4} }{
\Big(P^2_Si \Big)^{3/2} \Big(P^2_Ti \Big)^{1/2} }=
\frac{ (9i)^{4/4}(i)^{4/4}   }{ 
(4i)^{3/2} (\frac{3}{2}i)^{1/2}  }
=i\sqrt{\frac{27}{32}}.
\label{A221-loop}
\end{eqnarray}\\

A summary of this section is as follows.
We considered the reduced model with 
$A_{2,2}$ algebra, and expanded it around a background described by $A_{2,2}$ generators. 
We encountered many negative norm states, tachyonic modes, and a tadpole term if we use original action (\ref{action}). We then deformed the action with a forth-oder term, and the resulting action does not suffer from the above problems. After  gauge fixing of the deformed action, there are no quadratic terms of $A^{IJ}$, $S^{I\neq J}$, and gauge-fixing ghosts. The 1-loop determinant was then calculated in (\ref{A221-loop}) using eigenvalues $P^2$ and $\Gamma^{IJ}P_{IJ}$ of $T$, $S^{I=J}$, and fermions.  \\     

The fermionic symmetry (\ref{fersymm}) is no longer  a symmetry of the deformed action  (\ref{deformedaction}), but this action is invariant under the  new fermionic symmetry:
\begin{eqnarray}
\delta_{\chi} \hat{X}^I=i\bar{\chi}\Gamma^I\Psi, \quad
\delta_{\chi}\Psi=\frac{i}{6}\Gamma^{IJK}\left(
[ \hat{X}_{I}, \hat{X}_{J}, \hat{X}_{K}] - i\epsilon_{IJKL}\hat{X}^L
\right)\chi.
\label{defermsymn}
\end{eqnarray}        
Although the background (\ref{A22}) satisfies $\delta_{\chi}\Psi=0$, the bosonic and fermionic determinants do not completely cancel each other. Probably this implies an effects of curved spacetime. A similar effect has been reported for the  matrix model on the fuzzy sphere \cite{Imai:2003vr}.


\section{Nambu-Poisson Algebra}
In this section,  we use a Nambu-bracket:
$[ \; , \; ,\;]=-i\{ \;, \;, \; \}_{NP}$, where
\begin{eqnarray}
\{A,B,C\}_{NP}=\epsilon^{\mu\nu\rho} \frac{\partial A}{\partial x^{\mu}}\frac{\partial B}{\partial x^{\nu}}
\frac{\partial C}{\partial x^{\rho}}
\end{eqnarray}
with a signature $\eta^{\mu\nu}=diag(-1,-1,+1)$.
We take $p^{\mu}=x^{\mu}$ as the background, and expansion around the background is 
\begin{eqnarray}
\hat{X}^{\mu}&=&x^{\mu}+
\frac{1}{2}\epsilon^{\mu\nu\rho}b_{\mu\nu}(x), \\
\hat{X}^{4}&=&\phi(x), \\
\Psi&=&\psi(x). 
\end{eqnarray} 
$S^{(1)}$ now vanishes, and the quadratic part of the Lagrangian density comes from $S^{(2)}$ becomes
\begin{eqnarray}
\frac{1}{12}H^{\mu\nu\rho}H_{\mu\nu\rho}+\frac{1}{2}\partial^{\mu}\phi\partial_{\mu}\phi+\frac{i}{4}\bar{\psi}\Gamma^{\mu\nu}\epsilon_{\mu\nu\rho}\partial^{\rho}\psi
\end{eqnarray}
where $H_{\mu\nu\rho}= \partial_{\mu}b_{\nu\rho}+ \partial_{\nu}b_{\rho\mu}+ \partial_{\rho}b_{\mu\nu}$. From (\ref{gaugetransform}), the gauge transformations of these fluctuations become
\begin{eqnarray}
\delta_{\lambda}b_{\mu\nu}&=&i
 (\partial_{\mu}\lambda_{\nu}-\partial_{\nu}\lambda_{\mu})
+i\epsilon^{\alpha\beta\gamma}\partial_{\alpha}\lambda_{\beta}\partial_{\gamma}b_{\mu\nu},\\
\delta_{\lambda}\phi&=&i\epsilon_{\mu\nu\rho}\partial^{\mu}\lambda^{\nu}\partial^{\rho}\phi, \\
\delta_{\lambda}\psi&=&i\epsilon_{\mu\nu\rho}\partial^{\mu}\lambda^{\nu}\partial^{\rho}\psi.
\end{eqnarray} 
To calculate the 1-loop determinant, we need to fix a gauge of $b_{\mu\nu}$. For this purpose, we deform the action with a BRST exact term:
\begin{eqnarray}
\delta_BF=\delta_B\left[-\bar{c}^{\mu}(\alpha_{1}B_{\mu}+i\partial^{\nu}b_{\nu\mu})\right]
=-B^{\mu}(\alpha_{1}B_{\mu}+i\partial^{\nu}b_{\nu\mu})
+i\bar{c}^{\mu}\partial^{\nu}\delta_B b_{\nu\mu}
\label{bF}
\end{eqnarray}
where $c^{\mu}$ and $\bar{c}_{\mu}$ are the ghost and anti-ghost respectively, $B_{\mu}$ is the NL field, and $\alpha_{1}$ is a gauge parameter. 
BRST transformations of these fields are defined as 
\begin{eqnarray}
&&\delta_{B}b_{\mu\nu}=i(\partial_{\mu}c_{\nu}-\partial_{\nu}c_{\mu})
+i\epsilon^{\alpha\beta\gamma}\partial_{\alpha}c_{\beta}\partial_{\gamma}b_{\mu\nu},
\nonumber \\
&&\delta_{B}\phi=i\epsilon_{\mu\nu\rho}\partial^{\mu}c_{\nu}\partial_{\rho}\phi, \quad \delta_{B}\psi=i\epsilon_{\mu\nu\rho}\partial^{\mu}c_{\nu}\partial_{\rho}\psi,
\label{brstsec3}\\
&&\delta_{B}(\epsilon_{\mu\nu\rho}\partial^{\nu}c^{\rho})=
-i(\epsilon_{\alpha\beta\gamma}\partial^{\beta}c^{\gamma})\partial^{\alpha}(\epsilon_{\mu\nu\rho}\partial^{\nu}c^{\rho}),\quad 
\delta_B \bar{c}_{\mu}=B_{\mu}, \quad \delta_B B_{\mu}=0. \nonumber 
\end{eqnarray}
After integration of $B_{\mu}$ in (\ref{bF}), we have a gauge fixing term ${\cal L}_{gf}$ and ghost Lagrangian ${\cal L}_{gh}$:
\begin{eqnarray}
{\cal L}_{gf}&=&\frac{1}{4\alpha_{1}}(\partial^{\nu}b_{\nu\mu})(\partial_{\rho}b^{\rho\mu}),\\
{\cal L}_{gh}&=&-\bar{c}^{\mu}\partial^{\nu}[
\partial_{\nu}c_{\mu}-\partial_{\mu}c_{\nu}
-(\epsilon_{\alpha\beta\gamma}\partial^{\alpha}c^{\beta})\partial_{\gamma}b_{\mu\nu}
].
\end{eqnarray}
Taking $\alpha_{1}=\frac{1}{2}$, the quadratic part of the Lagrangian density becomes
\begin{eqnarray}
&& {\cal L}^{(2)} +{\cal L}_{gf} +{\cal L}_{gh} \;\rightarrow \nonumber \\
&&\frac{1}{4}(\partial^{\mu}b^{\nu\rho})(\partial_{\mu}b_{\nu\rho})
-\bar{c}^{\mu}\partial^{\nu}(\partial_{\nu}c_{\mu}-\partial_{\mu}c_{\nu})
+\frac{1}{2}\partial^{\mu}\phi\partial_{\mu}\phi
+\frac{i}{4}\bar{\psi}\Gamma^{\mu\nu}\epsilon_{\mu\nu\rho}\partial^{\rho}\psi.
\label{gaugefixedS2insec3}
\end{eqnarray} 

It is well known that  this is not the end of the gauge-fixing procedure. 
The ghost Lagrangian now has  gauge symmetry, which is a consequence of  
the gauge parameter $\lambda_{\mu}$ itself having  gauge symmetry.
We need to fix this gauge freedom by introducing another gauge fixing term for the ghost Lagrangian.
For this purpose, we again deform the Lagrangian by introducing another BRST-exact term:
\begin{eqnarray}
\delta_{B}G&=&\delta_{B}\left[
i\alpha_2(\partial^{\mu}\bar{c}_{\mu}) \pi-\alpha_3\bar{\rho}\pi+\alpha_4\bar{\beta}\rho
-i\alpha_5\bar{\beta}\partial^{\mu}c_{\mu}+\alpha_{6}\bar{\beta}\Box\sigma
\right] \nonumber \\
&=&i\alpha_2(\partial^{\mu}B_{\mu})\pi-i\alpha_5\bar{\beta}\partial^{\mu}(\delta_B c_{\mu})+\alpha_6\bar{\beta}\Box\beta
\nonumber \\
&&+\bar{\rho}(\alpha_3\rho-i\alpha_5\partial^{\mu}c_{\mu}+\alpha_6\Box\sigma)
+(\alpha_4\bar{\rho}-i\alpha_2\partial^{\mu}\bar{c}_{\mu})\rho.
\end{eqnarray}    
where $\alpha_2, ....,  \alpha_6$ are gauge parameters, and $\Box=\partial^{\mu}\partial_{\mu}$.
The assignment of ghost numbers to various fields in $\delta_BF$ and $\delta_BG$ is summarized by  
\begin{eqnarray}
\mbox{ghost number} &{}& \mbox{fields}  
\nonumber \\
2\quad &:& \beta \nonumber \\
1\quad &:& c_{\mu},  \rho, \sigma  \nonumber \\
0\quad &:& b_{\mu\nu}, \phi, \psi, B_{\mu},\pi \\
-1\quad &:& \bar{c}_{\mu}, \bar{\rho}, \nonumber \\
-2\quad &:& \bar{\beta} \nonumber 
\end{eqnarray}  
. BRST transformations are defined by
\begin{eqnarray}
\delta_B\bar{\beta}=\bar{\rho}, \quad 
\delta_B\bar{\rho}=0, \quad 
\delta_B\pi=\rho, \quad
\delta_B\rho=0, \quad
\delta_B \sigma=\beta, \quad 
\delta_B \beta=0 
\label{brstsec32}
\end{eqnarray}
together with (\ref{brstsec3}). 
We now consider the gauge boson part 
\begin{eqnarray}
\frac{1}{12}H^{\mu\nu\rho}H_{\mu\nu\rho}+\delta_BF+\delta_BG.
\end{eqnarray}
First, integrating out $B_{\mu}$ gives 
\begin{eqnarray}
\frac{1}{4\alpha_1}(\partial^{\nu}b_{\nu\mu}+\alpha_2\partial_{\mu}\pi)^2
=\frac{1}{4\alpha_1}(\partial^{\nu}b_{\mu\nu})(\partial_{\rho}b^{\rho\mu})
+\frac{\alpha_2^2}{4\alpha_1}\partial^{\mu}\pi\partial_{\mu}\pi.
\end{eqnarray}
Second, by integrating out $\rho$ and $\bar{\rho}$, we obtain
\begin{eqnarray}
\frac{\alpha_2\alpha_5}{\alpha_3+\alpha_4}(\partial^{\mu}\bar{c}_{\mu})\left(\partial^{\nu}c_{\nu}+i\frac{\alpha_6}{\alpha_5}\Box \sigma\right).
\end{eqnarray}
Third, we shift the ghost as $c_{\mu}+i\frac{\alpha_6}{\alpha_5}\partial_{\mu}\sigma \rightarrow c_{\mu}$, 
but this dose not change the ghost kinetic term.  Thus,  under a choice of gauge parameters
\begin{eqnarray}
\alpha_1=\frac{1}{2},\;\;  \alpha_2=1,\;\;
 \frac{\alpha_2\alpha_5}{\alpha_3+\alpha_4}=1,\;\; \alpha_6=-1,  
\end{eqnarray}  
the quadratic part of the gauge-fixing Lagrangian becomes
\begin{eqnarray}
\frac{1}{4}\partial^{\mu}b^{\nu\rho}\partial_{\mu}b_{\nu\rho}-\bar{c}^{\mu}\Box c_{\mu}
-\bar{\beta}\Box\beta+\frac{1}{2}\partial^{\mu}\pi\partial_{\mu}\pi.
\end{eqnarray}
The 1-loop determinants of $b_{\mu\nu}$ and ghosts are then
\begin{eqnarray}
\frac{  \det_{c\bar{c}} (i\Box ) }{\det_{b_{\mu\nu}}^{1/2} (i\Box ) \det_{\beta\bar{\beta}} (i\Box )  \det^{1/2}_{\phi} (i\Box ) }
=\frac{ \det^3(i\Box) }{ \det^{3/2}(i\Box) \det(i\Box) \det^{1/2}(i\Box) }=1
\end{eqnarray}
Therefore, the entire contribution from $b_{\mu\nu}$ is canceled by ghosts and ghosts for ghosts. The 1-loop determinant from the gauge field $b_{\mu\nu}$ turns out to be trivial in itself. This implies that $b_{\mu\nu}$ dose not have physical degrees of freedom.\\

Next we consider the 1-loop determinant from $\psi$. From 
\begin{eqnarray}
\left(\frac{1}{2}\Gamma^{\mu\nu}\epsilon_{\mu\nu\rho}\partial^{\rho}\right)
\left(\frac{1}{2}\Gamma^{\sigma\tau}\epsilon_{\sigma\tau\lambda}\partial^{\lambda}\right)
=-\Box,
\end{eqnarray}
the determinant becomes  $\det^{1/2}(i\Box)$. This is canceled by the 1-loop determinant $\det^{-1/2}(i\Box)$, which comes from $\phi$. 
Therefore, the whole 1-loop determinant equals to be 1: i.e., there is no vacuum energy. This is  as expected from supersymmetry in flat space.  \\

So far, we have considered the background of the reduced model $p^{\mu}=x^{\mu}$, but we may take a more general background $p^{\mu}=p^{\mu}(x)$. In this case, $P^{\mu\nu}=i\epsilon^{abc}\partial_{a}p^\mu\partial_{b}p^{\nu}\partial_{c}$. 
And $[P^{\mu\nu}, P^{\rho\sigma}]$ is generally non-vanishing. 
One can follow the discussion in section 2 and reach a gauge-fixed quadratic action (\ref{gaugefixedS2}). 
Next, writing  $\tilde{P}_\mu=\frac{1}{2}\epsilon_{\mu\nu\rho}P^{\nu\rho}$,
the action becomes
\begin{eqnarray}
\frac{1}{2}X^{\mu}(\tilde{P}^2\eta_{\mu\nu}-2[\tilde{P}_{\mu}, \tilde{P}_{\nu}]  )X^{\nu}
+\bar{c}_{\mu}( \eta^{\mu\nu}\tilde{P}^2- \tilde{P}^{\nu}\tilde{P}^{\mu})c_{\nu}
+\frac{1}{2}\phi\tilde{P}^2\phi
+\frac{1}{4}\bar{\Psi}\Gamma^{\mu\nu}\epsilon_{\mu\nu\rho}\tilde{P}^{\rho}\Psi
\end{eqnarray}
(We changed the sign of the ghost kinetic term in (\ref{gaugefixedS2}) with a suitable choice of the BRST-exact term.). 
If one rewrites $X^{\mu}$ as $b_{\mu\nu}=\epsilon_{\mu\nu\rho}X^{\rho}$, the obtained action is almost the same as (\ref{gaugefixedS2insec3}) except for the existence of terms such as  $[\tilde{P}, \tilde{P}]$. We can then repeat the procedure to introduce ghosts for ghosts and calculate the 1-loop determinant. 
The gauge boson and ghost Lagrangian with gauge fixing terms becomes       
 \begin{eqnarray}
\frac{1}{2}X^{\mu}(\tilde{P}^2\eta_{\mu\nu}-2[\tilde{P}_\mu, \tilde{P}_{\nu}])X^{\nu}
+\bar{c}^{\mu}(\tilde{P}^2\eta_{\mu\nu}+[ \tilde{P}_\mu, \tilde{P}_{\nu}])c^{\nu}
+\bar{\beta}\tilde{P}^2\beta+\frac{1}{2}\pi\tilde{P}^2\pi \nonumber \\
+\bar{c}^{\mu}[  \tilde{P}_{\mu}, \tilde{P}_{\nu} ](\tilde{P}^{\nu}\sigma)
-\pi\epsilon_{\mu\nu\rho}[  \tilde{P}_\mu, \tilde{P}_{\nu} ]X^{\rho}
+s^{(1)}_{\mu}X^{\mu}
\end{eqnarray}
where we included the tadpole term from $S^{(1)}$. This term does not affect the 1-loop determinant itself, but it shifts the vacuum energy. The term for mixing  between $\bar{c}^{\mu}$ and $\sigma$: $\bar{c}^{\mu}[  \tilde{P}_{\mu}, \tilde{P}_{\nu} ](\tilde{P}^{\nu}\sigma)$ begins to contribute from the 2-loop  through interaction vertexes, and thus, we ignore it at this stage. The 1-loop determinant from the gauge boson and ghosts sector is then
\begin{eqnarray}
\frac{ \det i( \tilde{P}^2\eta_{\mu\nu}+ [\tilde{P}_\mu, \tilde{P}_{\nu} ] )   }{
(\det i\tilde{P}^2)^{3/2}
\det^{1/2} i( \tilde{P}^2\eta_{\mu\nu}- 2[\tilde{P}_\mu, \tilde{P}_{\nu} ]
-\epsilon_{\mu\rho\sigma}[\tilde{P}^{\rho}, \tilde{P}^{\sigma} ]
\frac{1}{\tilde{P}^2}[\tilde{P}_{\kappa}, \tilde{P}_{\tau} ]\epsilon^{\kappa\tau\nu}
)  },
\end{eqnarray}   
and the vacuum energy shift due to the tadpole is
\begin{eqnarray}
\left\langle-s^{(1)}_{\mu}\left(   
\tilde{P}^2\eta_{\mu\nu}- 2[\tilde{P}_\mu, \tilde{P}_{\nu} ]
-\epsilon_{\mu\rho\sigma}[\tilde{P}^{\rho}, \tilde{P}^{\sigma} ]
\frac{1}{\tilde{P}^2}[\tilde{P}_{\kappa}, \tilde{P}_{\tau} ]\epsilon^{\kappa\tau\nu} 
\right)^{-1}s^{(1)}_{\nu} \right\rangle.
\end{eqnarray} 

The fermionic part of the 1-loop determinant also has the correction of $[\tilde{P}, \tilde{P}]$, and is calculated as  $\det^{1/2}i(\tilde{P}^2-\frac{1}{2}\Gamma^{\mu\nu}[\tilde{P}_\mu, \tilde{P}_{\nu} ])$. Adding the contribution of $\phi$, we obtain the whole result. Here we show the result in terms of vacuum energy:
\begin{eqnarray}
\frac{1}{2}\mbox{Tr}\log\left(
\eta_{\mu\nu}- 2[\tilde{P}_\mu, \tilde{P}_{\nu}] \frac{1}{\tilde{P}^2} 
-\epsilon_{\mu\rho\sigma}[\tilde{P}^{\rho}, \tilde{P}^{\sigma} ]
\frac{1}{\tilde{P}^2}[\tilde{P}_{\kappa}, \tilde{P}_{\tau} ]\epsilon^{\kappa\tau\nu}\frac{1}{\tilde{P}^2}\right)
\nonumber \\
-\mbox{Tr}\log\left(
\eta_{\mu\nu}+[\tilde{P}_\mu, \tilde{P}_{\nu}] \frac{1}{\tilde{P}^2}  \right)
-\frac{1}{4}\mbox{Tr}\log\left(
1- \frac{1}{2}\Gamma^{\mu\nu}[\tilde{P}_\mu, \tilde{P}_{\nu}] \frac{1}{\tilde{P}^2} \right)
\nonumber \\
-\left\langle s^{(1)}_{\mu}\left(   
\tilde{P}^2\eta_{\mu\nu}- 2[\tilde{P}_\mu, \tilde{P}_{\nu} ]
-\epsilon_{\mu\rho\sigma}[\tilde{P}^{\rho}, \tilde{P}^{\sigma} ]
\frac{1}{\tilde{P}^2}[\tilde{P}_{\kappa}, \tilde{P}_{\tau} ]\epsilon^{\kappa\tau\nu}
\right)^{-1}s^{(1)}_{\nu} \right\rangle.
\label{vacuenergysec4}
\end{eqnarray}  
Note that there is a supersymmetric cancellation of $\log \tilde{P}^2$, but terms due to the $[\tilde{P}, \tilde{P}]$ correction are non-vanishing. These corrections are due to deviation from flat commutative spacetime.
Here the  similarity to the matrix model on the fuzzy sphere is more evident than in the  previous section. \\

A summary of this section is as follows.
We considered Nambu-Poission algebra as the 3-algebra. In this case, the gauge symmetry becomes to  reducible symmetry. As a consequence, bosonic fluctuations can be written as one 2-form gauge potential and one scalar in three-dimension. To handle this reducible gauge symmetry, we introduced ghosts for ghosts employing the BRST gauge fixing procedure. The 1-loop determinant of the 2-form gauge potential turns out to be trivial itself if the background gives flat spacetime. This is consistent with an expectation that there are no physical degrees of freedom for the 2-form gauge field in three-dimensional spacetime. 1-loop determinants of the scalar boson and fermion cancel each other. The total 1-loop vacuum energy is then zero as expected from the supersymmetry in flat three-dimensional spacetime.
Next we considered backgrounds that give non-zero $[\tilde{P},\tilde{P}]$. In this case, supersymmetric cancellation is not perfect and the 1-loop vacuum energy has subleading remnants.\\

Finally, instead of using the Nambu-Poisson bracket, we may consider using the quantum Nambu-Poisson bracket. Formally it seems to be parallel, and one can reach the formula (\ref{vacuenergysec4}) for the 1-loop vacuum energy of a quantum background. 
However, it is pointed out in \cite{Nambu:1973qe} that 
$\tilde{P}^{\mu}$ for the quantum Nambu-bracket no longer satisfies the Leibniz rule. To find a set of eigenfunctions will be  a more difficult problem.


\section{Lorentzian 3-Algebra}
 In this section we use a Lorentzian 3-algebra as the 3-algebra. 
Generators of the algebra are denoted as $\{T^a\}=\{T^{-1}, T^0, T^{i}\}$. Here $T^i$ are generators of a Lie algebra, and satisfy $[T^i, T^j]=i{f^{ij}}_kT^k$, and $\langle T^i T^j\rangle=h^{ij}$.
The  Lorentz 3-algebra is defined by
\begin{eqnarray}
&&[T^{-1},T^a,T^b]=0, \quad 
[T^0, T^i, T^j]=i{f^{ij}}_kT^k, \nonumber \\
&&{}[T^i, T^j, T^k]=if^{ijk}T^{-1} \;\;(=i{f^{ij}}_lh^{lk}T^{-1}).
\label{lorentz3}
\end{eqnarray}
 with the inner product
\begin{eqnarray}
&&
\langle T^{-1}T^{-1}\rangle=0, \quad 
\langle T^{-1}T^{0}\rangle= -1, \quad
\langle T^{-1}T^{i}\rangle=0, \nonumber \\ 
&&
\langle T^{0}T^{0}\rangle=0, \quad 
\langle T^{0}T^{1}\rangle=0,\quad
\langle T^{i}T^{j}\rangle=h^{ij}.
\label{lorentz32}
\end{eqnarray}
We expand $\hat{X}^I$ and $\Psi$ as
\begin{eqnarray}
\hat{X}^I=X_{-1}T^{-1}+x^IT^0+X^I_iT^i, \quad 
\Psi=\Psi_{-1}T^{-1}+\Psi_0 T^0+\psi_iT^i. 
\label{expandsec3}
\end{eqnarray} 
Substituting these, the action $(\ref{action})$ becomes 
\begin{eqnarray}
S&=&-\frac{1}{4}({x}^I{x}_I)\mbox{Tr}[X^J, X^K][X_J, X_K]
-\frac{1}{2}({x}^I{x}_J)\mbox{Tr}[X^J, X^K][X_K, X_I] \nonumber \\
&&-\frac{1}{2}x_I\mbox{Tr} \;\bar{\psi}\Gamma^{IJ}[X_J,\psi]
-\frac{1}{2}\bar{\Psi}_0\mbox{Tr}\;\Gamma^{IJ}[X_I,X_J]\psi.
\label{actionsec5}
\end{eqnarray}
The gauge transformations of each $\Phi_{-1}, \Phi_0, \Phi_i$ are
\begin{eqnarray}
&&\delta_{\lambda} \Phi_i=i{f_{i}}^{jk}\lambda^{(1)}_j\Phi_k+\Phi_0\lambda^{(2)}_i, \\
&&\delta_{\lambda} \Phi_0=0, \\
&&\delta_{\lambda} \Phi_{-1}=\lambda^{(2)}_i\Phi^i.
\end{eqnarray}
These expressions imply that each $\Phi_{-1}$, $\Phi_{0}$  and $\Phi_{i}$  should be treated in different ways.   
$\Phi_{-1}$ does not appear in the action (\ref{actionsec5}); therefore the  integral $\int {\cal D}\Phi_{-1}$ factors out from the partition function (\ref{partitionfunc}). 
$\Phi_0$ does not have gauge transformation and is similar to coupling constants in a matrix model rather than the ''matrix field'' $\Phi_i$. 
Therefore, in contrast with the case for $\Phi_i$, it is  better not to decompose $\Phi_0$ to a background and fluctuations.   

Hence, the reduced model action has $SO(2,2)$ symmetry, and we may choose particular frames of $x^{I}$ to carry out the integral $\int {\cal D}x^I$,  using this symmetry. They are separated into sectors:
\begin{eqnarray}
\mbox{(I) }&:& x^{I}=(u,0,0,0)  \qquad \mbox{timelike}, \label{xtime}\\
\mbox{(I{}I)}&:& x^{I}=(0,0,0,u) \qquad  \mbox{spacelike}, \label{xspace}\\ 
\mbox{(I{}I{}I)}&:& x^{I}=(u,0,0,\pm u) \qquad \mbox{null} \label{xnull}
\end{eqnarray} 
In each sector, the $x^I$ integral reduces to 
\begin{eqnarray}
\int {\cal D}x^I\;\rightarrow \; (\mbox{volume factor})\times \int^{\infty}_{0}u^3du.
\end{eqnarray}

In the following, we consider partition functions of these regions separately.

\subsection{ Case (I)}
We choose the frame (\ref{xtime}). The action (\ref{actionsec5}) becomes
\begin{eqnarray}
S&\rightarrow&
\frac{1}{4}u^2\mbox{Tr}[X^i,X^j][X_i,X_j]-\frac{1}{2}u\mbox{Tr}\bar{\psi}\Gamma^{1i}[X_{i},\psi]
\nonumber \\
&&-\frac{1}{4}\mbox{Tr}\bar{\Psi}_0\Gamma^{IJ}[X_I,X_J]-\frac{1}{4}\mbox{Tr}\bar{\psi}\Gamma^{IJ}[X_I,X_J]\Psi_0
\end{eqnarray}
where $i,j\neq 1$.
We can eliminate  $u$-dependence in the action with a  rescaling: 
\begin{eqnarray}
X\rightarrow u^{-\frac{1}{2}}X, \psi\rightarrow u^{-\frac{1}{4}}\psi, \Psi_0\rightarrow u^{\frac{3}{4}}\Psi_0.
\label{rescale}
\end{eqnarray} 
To see the effect of $\Psi_0$, we integrate out $\psi$ first. The resulting action is
\begin{eqnarray}
\frac{1}{4}\mbox{Tr}[X^i,X^j][X_i,X_j]-\frac{1}{2}\mbox{Tr}_{\psi}\log  [C^{-1}\Gamma^{1j}(adX_j)] 
\nonumber \\
-\frac{1}{8}\mbox{Tr}\bar{\Psi}_0\Gamma^{IJ}[X_I,X_J]\frac{1}{\Gamma^{1j}(adX_j)}[X_K,X_L]\Gamma^{KL}\Psi_0. 
\end{eqnarray}
where $adX_j=[X_j,\;\;]$. We now recall that $\Psi_0$ is a two-component real spinor, and thus, the integration of $\Psi_0$ is easily done.
Thus, the partition function after the $\Psi_0$ integral, up to a volume factor, is
\begin{eqnarray}
&&\hspace{-15mm}
Z= 
\int^{\infty}_{0}duu^{\frac{3}{2}(1-n_g)} \int{\cal D}X 
\det{} ^{1/2}_{\psi} [ iC^{-1}\Gamma^{1j}(adX_j)]
\nonumber \\ 
&&\frac{-i}{8}\epsilon_{\alpha\beta}\left(
\mbox{Tr}C^{-1}\Gamma^{IJ}[X_I,X_J]\frac{1}{\Gamma^{1j}(adX_j)}[X_K,X_L]\Gamma^{KL} 
\right)_{\alpha\beta} 
e^{i\frac{1}{4}\mbox{Tr}[X^i,X^j][X_i,X_j]}.
\end{eqnarray}   
 The factor $u^{\frac{3}{2}(1-n_g)}$ is the result of the rescalling (\ref{rescale}), where $n_g$ is the number of Lie algebra generators.
The $u$-integral then diverges as $u\rightarrow 0$. 
Another divergence comes from the $X^1$ integral, 
because $X^1$ appears in the factor $\Gamma^{IJ}[X_I,X_J]$ only, 
 and thus, there is  no convergent factor for $X^1$. 
One way of thinking may be that we integrate $X^i$ while $u$ and $X^1$ are fixed.  
Under some fixed value of $u$ and $X^i$, we can decompose $X^i$ as $X^i=p^i+a^i$, where $p^i$ is a background and $a^i$ is a fluctuation around it.  
Calculation of this 1-loop determinant is the same as that of 
 the IKKT type matrix model with (1+2)-dimensional target space. 
After introducing the gauge ghost, we obtain the  contribution
in terms of the effective action as calculated in \cite{Ishibashi:1996xs}
\!\footnote{There is also a tadpole contribution, if it exists.} 
\begin{eqnarray}
 \frac{1}{2}\mbox{Tr}\log (P^2\eta_{ij}+2[P_{i},P_{j}])-\mbox{Tr}\log P^2
-\frac{1}{2}\mbox{Tr}\log\left( P^2+\frac{1}{4}\Gamma^{ij}[P_i,P_j] \right).
\label{IKKT1loop}
\end{eqnarray}

Thus, the leading contributions of the fluctuation $\log P^2$ term  cancel each other. 
In addition to the non-vanishing subleading contributions in (\ref{IKKT1loop}), here we have another source of a contribution to the 1-loop effective action
that comes from the $\Psi_0$ integral: 
\begin{eqnarray}
-\log \epsilon^{\alpha\beta}\left(\mbox{Tr}C^{-1}\Gamma^{i1}(P_iX_1)\frac{1}{\Gamma^{1j}P_j}(P_kX_1)\Gamma^{k1}\right)_{\alpha\beta}. 
\end{eqnarray}  \\

The aspect of the spacelike  case (I{}I) is almost the same except for some changes of signs. Therefore we next consider the null case (I{}I{}I).

\subsection{ Case (I{}I{}I)}
The vector $x^I$ is null in the case (I{}I{}I). These are two cases of the choice of  the vector $x^{I}=(u,0,0,\pm u)$. Here we discuss in the case of the $''+''$ sign. The other choice gives a similar result. Substituting the expression of $x^I$, the reduced model action becomes 
\begin{eqnarray}
S\;\rightarrow\;
\frac{1}{2}u^2\mbox{Tr}[Z,X_i][Z,X^i]
+\frac{1}{2}u\mbox{Tr}\bar{\psi}\Gamma^{14}[Z,\psi]
+\frac{1}{2}u\mbox{Tr}\bar{\psi}\Gamma^{+}\Gamma^i[X_i,\psi]
\nonumber \\
-\frac{1}{2}\bar{\Psi}_0(\mbox{Tr}\Gamma^{IJ}[X_I,X_J]\psi)
-\frac{1}{2}(\mbox{Tr}\psi\Gamma^{IJ}[X_I,X_J])\Psi_0,
\end{eqnarray}     
where 
$i=2,3$, $Z=X^1-X^4$ and $\Gamma^{+}=\Gamma^{1}+\Gamma^{4}$. 
Similar to case (I) (and (I{}I)), we employ rescaling (\ref{rescale}), 
and the integral of fermions $\psi$ and $\Psi_0$.  
This gives
\begin{eqnarray}
&&\hspace{-15mm}
Z= 
\int^{\infty}_{0}duu^{\frac{3}{2}(1-n_g)}\int{\cal D}Z \int{\cal D}W\int{\cal D}X^i 
\det{} ^{1/2}_{\psi} [ iC^{-1} ( \Gamma^{14}(adZ)+ \Gamma^{+i}(adX_j)   ]
\nonumber \\ 
&&\hspace{-15mm}
\frac{-i}{8}\epsilon_{\alpha\beta}\left(
\mbox{Tr}C^{-1}\Gamma^{IJ}[X_I,X_J]\frac{1}{ \Gamma^{+i}(adX_i) + \Gamma^{14}(adZ)  }[X_K,X_L]\Gamma^{KL} 
\right)_{\alpha\beta} 
e^{i\frac{1}{2}\mbox{Tr}[Z,X_i][Z,X^i]}
\label{intparfunc2}
\end{eqnarray}    
up to a volume factor, where $W \equiv X^1+X^4$ is included in $\Gamma^{IJ}[X_I,X_J]$  terms. 

Using the explicit representation of gamma matrices in the appendix, we  calculate
\begin{eqnarray}
&&A=\Gamma^{14}(adZ)+\Gamma^+\Gamma^i(adX_i) 
\nonumber \\
&&\rightarrow\left( \begin{array}{cc}
 adX_2+adX_3 & adX_2+adX_3-adZ \\
 -(adX_2+adX_3+adZ) & -(adX_2+adX_3) 
\end{array}\right),
\end{eqnarray}  
where the Weyl projection is used to obtain the last line. 
The determinant of $A$ for spinor indexes then gives $det_{(spinor)} A= -(adZ)^2$.  
We may fix the value of $u$, $X^1, X^4$, and regard that $X^i$ (the part that satisfies $[Z, X^i]\neq 0$ ) is the fluctuation to be integrated. 
The Gauss-Fresnel integral of $X^i$ gives a  bosonic 1-loop determinant $\det^{2/2} [i(adZ)^2]$. 
Combining this with the fermion integral $\det^{1/2} iA$, 
we obtain a factor $\det^{-1/2} [i(adZ)^2]$. 
We can diagonalize $Z$ as a gauge fixing. Thus, $Z=diag(z_1,z_2, .... ,z_i, .... , z_N )$ and
$(adZ)_{ij}=z_i-z_j$. We have the Vandermonde determinant $\prod_{i\neq j}(z_i-z_j)^2$ in the partition function after the diagonalization of $Z$, and this determinant cancels the contribution $\det^{-1/2} [(adZ)^2]$ from the integration of $X^i$ and $\psi$.   

On the other hand,  the result of the $\Psi_0$ integral (second line in (\ref{intparfunc2})) 
gives a  contribution to the 1-loop effective action:
\begin{eqnarray}
-\log \epsilon^{\alpha\beta}
\left(
\frac{1}{4}\mbox{Tr}C^{-1}\Gamma^{14}[Z,W] \frac{1}{adZ} [Z,W]
\right)_{\alpha\beta}. 
\end{eqnarray} 
   \\

A summary of this section is as follows.
We used a simple Lorentzian 3-algebra. According to SO(2,2) symmetry, the path integral can be separated into sectors in which $x^I$ is timelike, spacelike, and a null vector respectively. In the case that $x^I$ is timelike (or spacelike), the  integral over $u (=|x^I|)$ and $X^1$ (or $X^4$) gives divergences, which may indicate that the reduced model prefers the configuration $u=\infty$ and $X^1=0$ (or $X^4=0$). 
We fixed these by hand and considered the integration over fluctuations around these configurations. Calculation of the 1-loop determinant is almost the same as in the case of the three-dimensional super-Yang-Mills matrix model, except for a new contribution comes from the $\Psi_0$ integral. This contribution has a  non-vanishing effect on the 1-loop vacuum energy. 

Next we considered  the case that $x^I$ is a null vector. The integral over $X^1, X^4$ and $u$ gives divergences. We fixed these value by hand, and considered to $X^{i=2,3}$ and $\psi$ as fluctuations to be integrated. In this case, the 1-loop determinant of these bosonic and fermionic fluctuations cancel each other, after including a factor due to diagonalizing $Z$ that corresponds to the gauge ghost determinant. In addition, there is a term from the $\Psi_0$ integral that makes a  non-vanishing contribution to the 1-loop vacuum energy.


\section{Summary and Discussion}
In this paper, we considered a supersymmetric reduced model with 3-algebraic structure, and several examples of 3-algebra were used to calculate 1-loop determinants around backgrounds that describe (fuzzy) spacetimes. 
Although we can have a formula of the 1-loop determinant  like (\ref{1-loop}), without further specification of the 3-algebra, it is neither useful nor accurate. 
We found that the reason for this is that the behavior of various fluctuations
 differ depending  on the choice 3-algebra.
There are modes that do not appear in the quadratic part of action (section 3).
Or, each of these modes can have different gauge transformation property;
 in particular, some of them do not transform (section 5). 
Moreover, there are cases that the gauge transformation becomes reducible (section 4).  
This observation suggests that we need to classify the 3-algebra before constructing  a generic formula for the quantum correction.    
The study of  such a systematic procedure is a future problem.
Among the  3-algebras that we investigated, the Lorentzian 3-algebra is closest to  IKKT type matrix models. 
From this view point, investigating this class of 3-algebra appears interesting.  
Several  realizations of 3-algebras in this class have been found \cite{Ho:2009nk} and a more formal argument has been developed \cite{Chu:2010nj}.     
Interesting results are expected using these algebras. 

The 1-loop determinants evaluated in this paper can be identified with 1-loop vacuum energies of corresponding (fuzzy) spacetimes. We found non-vanishing results even in the supersymmetric model. These results can be used for stability analysis of various background spacetimes of the reduced model. Including the evaluation of the higher loop effect, this is one future direction with a viewpoint similar to that of \cite{Imai:2003jb}. In the case of Lorentzian algebra, we found that the vacuum energy receives a new effect originating from the $\Psi_0$ integral, in addition to the known results of the IKKT type matrix model. Therefore, we expect that the effects from the M-theory direction give a correction to the stability analysis done by the IKKT type matrix model.\\

Finally, we point out that our analysis is applicable to the model 
that  is the dimensional reduction of the BLG model to zero dimension, discussed in \cite{Sato:2010ca}. This model relates to a supermembrane in eleven-dimension within a low derivative approximation. 
The action is 
\begin{eqnarray}
S&=&\Bigg\langle
\frac{1}{12}[X^I,X^J,X^K]^2+\frac{1}{2}(A_{\mu ab}[T^a,T^b,X^I])^2
-\frac{1}{3}\epsilon^{\mu\nu\lambda} A_{\mu ab}A_{\nu cd}A_{\lambda ef}[T^a,T^c,T^d][T^b,T^e,T^f]
\nonumber \\
&&-\frac{i}{2}\bar{\Psi}\Gamma^{\mu}A_{\mu ab}[T^a,T^b,\Psi]
+\frac{i}{4}\bar{\Psi}\Gamma^{IJ}[X_I,X_J,\Psi] \Bigg\rangle.
\end{eqnarray}
The main difference from the model in this paper is the existence of the Chern-Simon type gauge boson $A_{ab}^{\mu}$. 
Now let us consider the relevant part to the 1-loop determinant from the quadratic term of $A_{ab}^{\mu}$. 
In the case of  Nambu-Poisson algebra, for example, it can be written as 
\begin{eqnarray}
\left\langle -\frac{1}{2} 
(\epsilon^{ijk}\partial_i a_{j}^{\mu} )^2 \right\rangle.
\end{eqnarray}  
where $a_{j}^{\mu}=A_{ab}^{\mu}T^a\partial_{j}T^b$. 
Namely, they are field strengths of  the three gauge bosons labeled by $\mu=1,2,3$. Thus, after introducing the gauge fixing ghost for them, they gives the contribution $\det^{-3/2}(i\Box)$.      
As discussed in section 4, bosons $X^I$  contribute $\det^{-(D-3)/2}(i\Box)$ after gauge fixing, now $D=8$. Therefore, the total bosonic contribution becomes $\det^{-8/2}(i\Box)$ and it is canceled by the 1-loop determinant from the  fermion, as expected from supersymmetry in flat three-dimension.


\section*{Acknowledgments}

Author would like to thank
Pei-Ming Ho, 
Kazuyuki Furuuchi,
and Tomohisa Takimi
for discussions and comments.


\appendix
\section{$SO(2,2)$ Gamma Matrix and Majorana-Weyl Fermion}
We consider the $SO(2,2)$ Gamma matrix:
\begin{eqnarray}
\Gamma^I\Gamma^J+\Gamma^J\Gamma^I=2\eta^{IJ}, \quad 
\eta^{IJ}=diag(-1,-1,+1,+1).
\label{22gamma}
\end{eqnarray}
An explicit representation 
is given by
\begin{eqnarray}
\Gamma^1=\left(\begin{array}{cc}
0&-\sigma^1 \\
\sigma^1&0
\end{array}\right), 
\Gamma^2=\left(\begin{array}{cc}
0&-\sigma^3 \\
\sigma^3&0
\end{array}\right), 
\Gamma^3=\left(\begin{array}{cc}
0&-i\sigma^2 \\
i\sigma^2&0
\end{array}\right), 
\Gamma^4=\left(\begin{array}{cc}
0&\bf{1} \\
\bf{1}&0
\end{array}\right),
\end{eqnarray}
where $\sigma^i$ are Pauli matrices. The charge conjugation matrix is defined by the property $C^{-1}\Gamma^IC=+(\Gamma^I)^T$. 
A representation of such $C$ is  
\begin{eqnarray}
C=\left(\begin{array}{cc} 
-i\sigma^2& 0\\ 0 & -i\sigma^2
\end{array}\right).
\end{eqnarray}
Here $C^2=-1$ and $C^T=-C$.
In this representation, a four-component Majorana fermion can be written as two real two-component spinors as 
\begin{eqnarray}
\Psi_M=\left(\begin{array}{c}\chi_1 \\ i\chi_2\end{array}\right), \quad 
(\mbox{$\chi_1, \chi_2$: real}).
\label{majofermi}
\end{eqnarray} 
The Dirac conjugate $\bar{\Psi}_M$ can be written as
\begin{eqnarray}
\bar{\Psi}_M=\Psi_M^TC^{-1}.
\end{eqnarray}
On the other hand, $\Gamma_5$ becomes
\begin{eqnarray}
\Gamma_5=\Gamma_{1}\Gamma_{2}\Gamma_{3}\Gamma_{4}=
\left( \begin{array}{cc}
-\bf{1} & 0 \\ 0 & \bf{1}
\end{array}\right).
\end{eqnarray}
The Weyl fermion is defined by the projection condition $\Gamma_5\Psi_W=\pm\Psi_W$. Combining this with (\ref{majofermi}), the Majorana-Weyl fermion that satisfies  the Weyl condition $\Gamma_5\Psi_{MW}=-\Psi_{MW}$ becomes 
\begin{eqnarray} 
\Psi_{MW}=\left(\begin{array}{c} \chi_1 \\ 0 \end{array}\right).
\end{eqnarray}

For a more general introduction of the $SO(t,s)$ Gamma matrix and fermion, see \cite{Furuuchi:2009ax} and the references therein.

\newpage


}\end{document}